\documentclass[11pt]{article}
\usepackage{epsfig,color,a4wide}
\usepackage[margin=10pt,font=small]{caption}

\begin{document}

\renewcommand\floatpagefraction{.99}
\renewcommand\topfraction{.99}
\renewcommand\bottomfraction{.99}
\renewcommand\textfraction{.01}   
\setcounter{totalnumber}{50}
\setcounter{topnumber}{50}
\setcounter{bottomnumber}{50}

\renewcommand\theequation{\thesection.\arabic{equation}} 
\renewcommand\theequation{\arabic{equation}}

\parindent 0pt
\parskip 4pt
\newcommand\void[1]{}

\newcommand{\be}{\begin{equation}}
\newcommand{\ee}{\end{equation}}
\def\ds{\displaystyle}
\hbox{~}

\begin{flushright}  {~} \\[-12mm]
{\sf hep-th/0612203}\\[1mm]
{\sf KCL-MTH-06-17}\\[1mm]
\end{flushright} 

\vskip 14mm
{\LARGE\bf 
A Renormalisation group for TCSA
}\\[15mm] 
{\large 
Giovanni Feverati,\,
Kevin Graham,\,
Paul~A.~Pearce,\, Gabor Zs.~T\'oth, and
\underline{G\'erard Watts}}\footnote{
Talk given by G.M.T. Watts at the workshop {\em ``Integrable Models and Applications:
from Strings to Condensed Matter''}, 
Santiago de Compostela, Spain, 12-16 September 2005 
}
\\
{\parindent 15pt\parskip 0pt

{\it\small 

LAPTH, 9 Chemin de Bellevue, BP 110 74941 Annecy le Vieux Cedex --
France

Institut f\"ur Theoretische Physik, Freie Universit\"at Berlin,
Arnimallee 14, 14195 Berlin -- Germany

Dept.\ of Mathematics \& Statistics,
University of Melbourne
Parkville, VIC, 3010 -- Australia

Dept.\ of Theoretical Physics, 
ELTE,
P\'azm\'any P\'eter S\'et\'any 1/A,
Budapest 1117 -- Hungary

Dept.\ of Mathematics, King's College London,
Strand, London WC2R 2LS -- UK

}
}
\vskip 10mm

{\bf Abstract}

We discuss the errors introduced by level truncation in the
study of boundary renormalisation group flows by the Truncated
Conformal Space Approach. We show that the TCSA results can have the
qualitative form of a sequence of RG flows between different conformal
boundary conditions. 
In the case of a perturbation by the field $\phi_{(13)}$, we propose a
renormalisation group equation for the coupling constant which 
predicts a fixed point at a finite value of the TCSA coupling constant
and we compare the predictions with data obtained using TBA equations.

\section{Perturbed boundary conformal field theory}

The Truncated Conformal Space Approach (TCSA) is a tool to study
finite size effects or RG flows in perturbed conformal field theory
\cite{YZam1,BTBA}.  
Here we consider boundary RG flows where conformal
invariance of a system with a boundary is broken by a coupling to a
boundary field. 
\begin{equation} 
\delta S = \lambda \int \phi(x) dx 
\;.
\label{eq1}
\end{equation}
These are easier to study than bulk flows in many ways -- for a
unitary theory, the UV and IR fixed points must be conformal boundary
conditions which are well understood, and the boundary entropy $g$
must decrease along the flow \cite{AL91,Kon1}.

\begin{figure}[h]
\begin{center}
\input{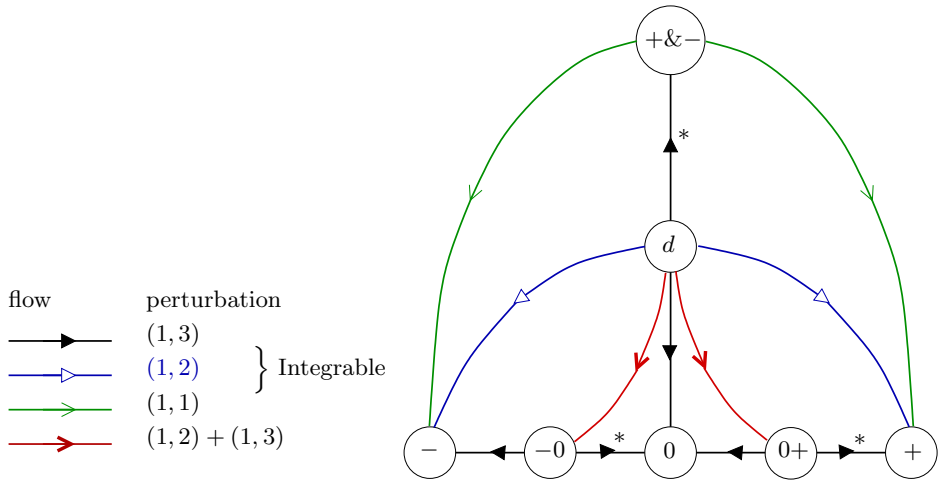}
\end{center}
\caption{The space of boundary flows in the tricritical Ising model}
\label{fig1}
\end{figure}

An example of a problem is to study the space of flows in the tri-critical
Ising model. 
The tri-critical Ising model is a unitary conformal field theory with central
charge $7/10$ and contains 6 representations of the Virasoro algebra
and correspondingly 6 bulk primary fields.
The conformal weights and labelling of the
Virasoro representations are
\[
\begin{array}{c|cccccc}
\hbox{\it label} & (11) & (21) & (31) & (12) & (13) & (22) \\[2mm]
\hline&&&&&&\\[-4mm]
\hbox{\it conformal weight} & 0 & \frac 7{16} & \frac 32 & \frac1{10}
& \frac{3}{5} & \frac 3{80} 
\end{array}
\]
For this model, there are thus 6 fundamental conformal boundary
conditions corresponding to the 6 primary fields \cite{Cardy1}.
It is the continuum limit of an RSOS lattice model with heights taking
integer values 1 to 4, or a spin model where the spins take values
$-$, $0$ and $+$.
One can realise the boundary conditions in terms of restrictions on
the values that the spins on the edge can take.
Furthermore, each conformal boundary condition supports boundary
fields organised into representations of the Virasoro algebra, the
representations given by the fusion rules.
For the 6 fundamental boundary conditions the values of $g$ and the
conformal families of boundary fields are
\[
\begin{array}{r|c|c|c|c|c|c}
\hbox{\it spins} & (-) & (0) & (+) & (-0) & (0+) & (-0+) \\[4mm]
\hbox{\it labels} & (11) & (21) & (31) & (12) & (13) & (22) \\[2mm]
\hline&&&&&&
\\[-4mm]
g & 0.5127 & 0.725 & 0.5127 & 0.8296 & 0.8296 & 1.173 \\[4mm]
\hbox{\it boundary fields} & (11) & (11),(31) & (11) &
(11), {\color{red}(13)} 
& 
(11), {\color{red}(13)} 
& 
(11), {\color{red}(13)},{\color{red}(12)},(31) 
\\
\end{array}
\]
The condition for a boundary field to be relevant (i.e. to generate a
boundary flow) is that its weight be less than 1 and in this model
these are just the 
fields labelled $(12)$ and $(13)$ together with the field $(11)$ of
weight zero.
The boundary RG flows of the tricritical Ising model have been well
studied and the global picture in figure \ref{fig1} first proposed by
Affleck \cite{Affl1}.
This can be checked using TCSA, and many of the flows can be checked
by other methods as well, in particular the flows marked $*$ agree
with perturbtion theory \cite{RRS} and the integrable flows generated
by the perturbation $\phi_{(13)}$ have been studied using TBA
equations derived from a 
lattice approach \cite{FPR1}.

\section{The Truncated Conformal Space Approach}

A strip with the perturbation (\ref{eq1})
along a boundary is equivalent to a system with a perturbed\newline
Hamiltonian
\be 
  H = H_0 + \lambda\phi(0) 
\;.
\ee
If the strip has conformal boundary conditions $(a)$ 
and $(b)$  on its two edges
then the Hilbert space on which the Hamiltonian acts is given by the fusion rule
\be
{\cal H} = \oplus_c N_{ab}{}^c {\cal H}_c
\;.
\ee
If we take $(a)$  to correspond to the identity operator then the
Hilbert space is a single representation of the Virasoro algebra.
\[{\cal H = H}_b\]
It is convenient to map the strip to the upper half plane and consider
the operator 
\be
 \frac R{\pi}H = (L_0-\frac{c}{24}) +
 \lambda(\frac R\pi)^{1-h}\phi(1)
\;.
\ee
The matrix elements of this operator can be
calculated exactly.
This gives an infinite matrix, and the idea of TCSA is to truncate the
Hilbert space to states with energy less than or equal to $(N+h)$. 
One can then diagonalise the resulting matrices and investigate the
spectrum and other properties of the truncated system.

If we perturb the $(b)$  boundary so that in the IR it flows to a new
boundary $(b')$  then the spectrum of the Hamiltonian will interpolate
that of the UV and IR boundaries.
Since
\be
H_0 = \left(\frac{\pi}{R}\right)(L_0 - \frac c{24})
\;,
\ee
then in the case that the Hilbert space consists of a single
representation of the Virasoro algebra, in all but the vacuum
representation the normalised energy gaps 
\be
\Delta_i = \frac{E_i - E_0}{E_1 - E_0}
\ee
will be integers at the UV and IR fixed points with 
multiplicities that are given by the characters
of the two representations $(b)$  and $(b')$.
In the vacuum representation since the first excited state has $L_0$
eigenvalue 2, the normalised gaps will be half-integers.
For the tri-critical Ising model these multiplicities are given in
table \ref{tab1}.
\begin{table}[htb]
\[
{\renewcommand{\arraystretch}{1.1}
\begin{array}{rr|rrrrrrrrrrrrr}
&& 
       ~~0 & ~1 & ~2 & ~3 & ~4 & ~5 & ~6 &  7 &  8 & 9 & 10 & \cdots\\
\hline
\Delta_i 
&(12) & 1 & 1 & 1 & 2 & 3 & 4 & 6 &  8 & 11 & 14 & 19 & \cdots \\
&(13) & 1 & 1 & 2 & 2 & 4 & 5 & 7 &  9 & 13 & 16 & 22 & \cdots \\
&(21) & 1 & 1 & 1 & 2 & 3 & 4 & 6 &  8 & 10 & 14 & 18 & \cdots \\
&(31) & 1 & 1 & 2 & 2 & 3 & 4 & 6 &  7 & 10 & 12 & 16 & \cdots \\
&(22) & 1 & 1 & 2 & 3 & 4 & 6 & 8 & 11 & 15 & 20 & 26 & \cdots \\
\hline
2\Delta_i & 
(11) & 1 & 0 & 1 & 1 & 2 & 2 & 4 &  4 &  7 &  8 & 12 & \cdots \\
\hline
\end{array}
}
\]
\caption{Multiplicities of low lying states in the Virasoro
  representations entering the tri-critical Ising model}
\label{tab1}
\end{table}

\begin{figure}[hbt]
\begin{center}
\vspace{3mm}
\begin{tabular}{cc}
\epsfxsize=7cm %12cm
\epsfbox{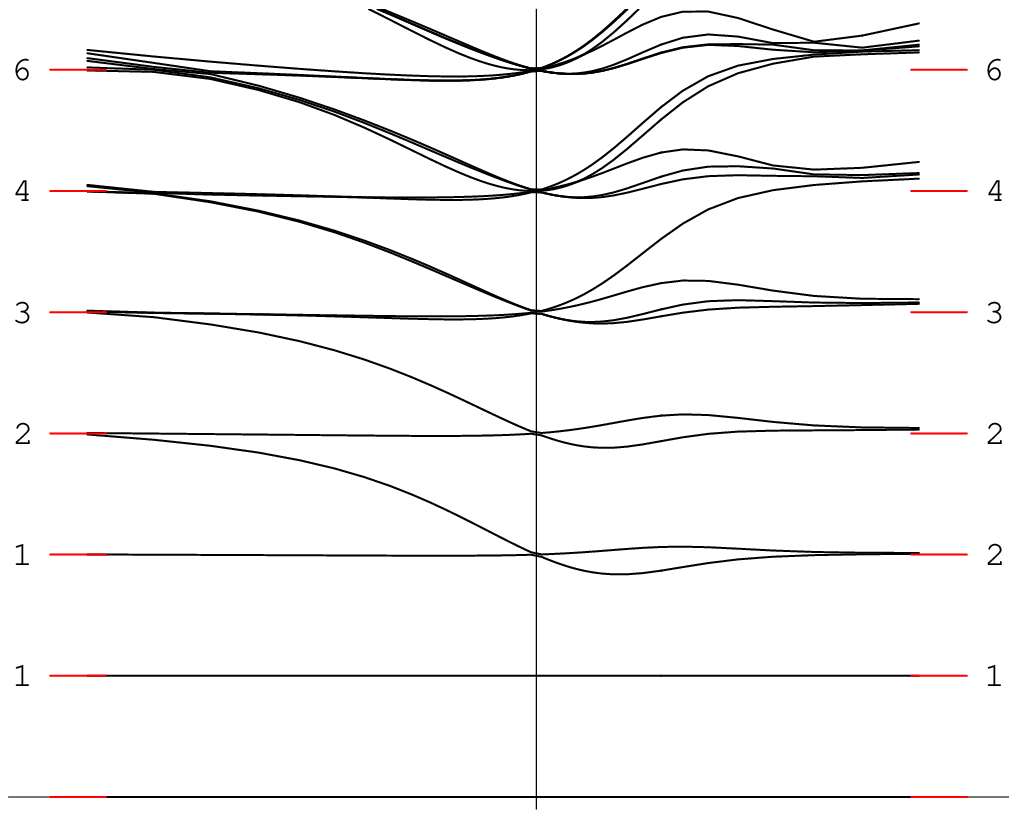}
&
\epsfxsize=7cm %12cm
\epsfbox{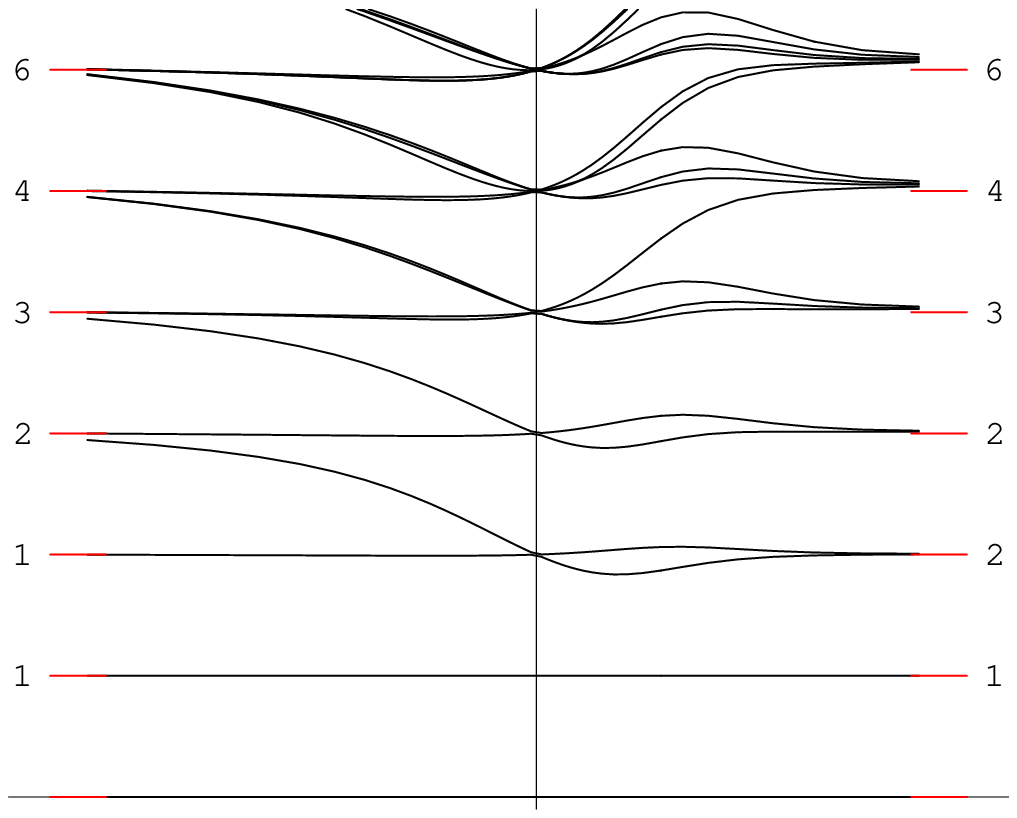}
\\
Level 10: 82 states
&
Level 16: 410 states
\end{tabular}
\end{center}
\vspace{-5mm}
\caption{The normalised energy gaps for the pertubation of the strip
  with boundary conditions $(11)$ and $(13)$ by the field
  $\phi_{(13)}$ at two different truncation levels. The multiplicities
  of the $(31)$ and $(21)$ representations are shown on the right and
  left for comparison}
\label{fig2}
\end{figure}

As a concrete example, consider the flows away from the boundary
condition (13) generated by the field $\phi_{(13)}$.
We can expect that the boundary condition flows for one sign of the coupling to
the character $(31)$  and for the other to $(21)$
In figure \ref{fig2} we show the normalised energy gaps $\Delta_i$ for
the perturbation of the theory on a strip with boundary conditions
$(11)$ and $(13)$ by the field $\phi_{13}$ on the $(13)$ boundary.
For zero coupling the multiplicities are those of the representation
$(13)$. For positive coupling they reorganise themselves approximately
into the multiplicities of the $(31)$ representation and for negative
coupling into those of the $(21)$ representation. in agreement with
perturbative and TBA calculations. Furthermore the accuracy with which
this reorganisation occurs increase with increasing truncation level
--- here from 82 states with $N=10$ to $410$ states with $N=16$.

\begin{figure}[h]
\begin{center}
\epsfxsize=10cm %14cm
\epsfbox{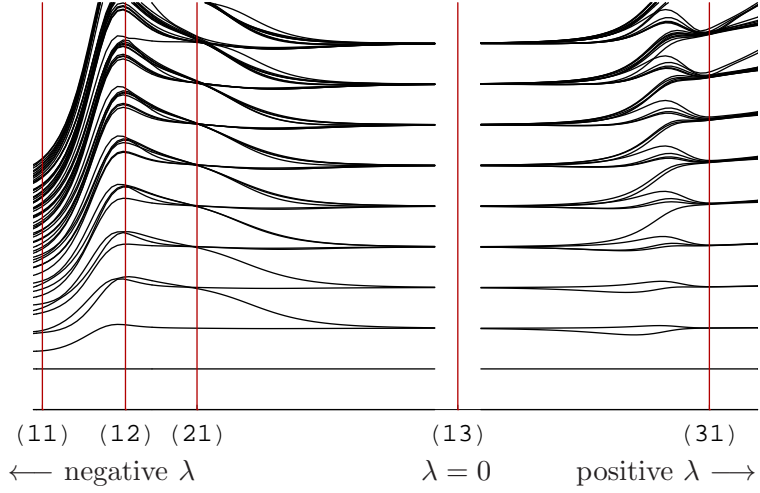}\\[-2mm]
\centerline{%
\hbox{%
$\longleftarrow$ negative $\lambda$ 
\hspace{2.876truecm}$\lambda=0$
\hspace{1truecm}positive $\lambda$ $\longrightarrow$}}
\end{center}
\vspace{-5mm}
\caption{The normalised gaps for the perturbation of the strip 
  with boundary conditions $(11)$ and $(13)$ by the field
  $\phi_{(13)}$ with positive and negative coupling on a logarithmic
  scale. The positions of the approximate fixed points $(31)$, $(21)$,
  $(12)$ and $(11)$ are indicated by vertical lines }
\label{fig3}
\end{figure}

\begin{figure}[h]
\begin{center}
\begin{tabular}{c}
\epsfxsize=12cm
\epsfbox{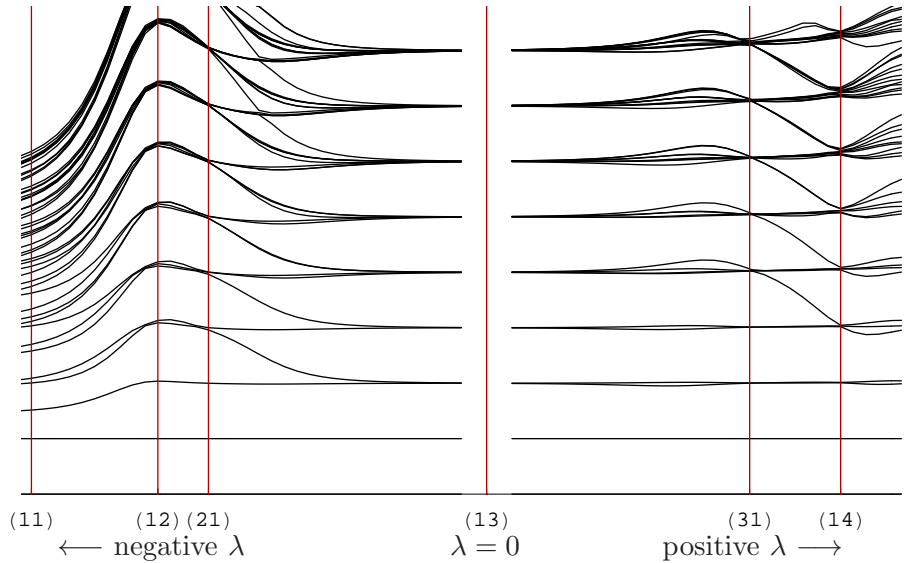}\\[-5mm]
\hbox{%
$\longleftarrow$ negative $\lambda$ 
\hspace{2.6truecm}$\lambda=0$
\hspace{1.765truecm}positive $\lambda$ $\longrightarrow$}\\[3mm]
\end{tabular}
\vspace{-5mm}
\caption{The normalised gaps for the perturbation of the strip 
  with boundary conditions $(11)$ and $(13)$ in the model $M_{6,7}$ by
  the field 
  $\phi_{(13)}$ with negative and positive 
  couplings each on a logarithmic
  scale at truncation level 16. The positions of the approximate fixed
  points are indicated by vertical lines. 
} 
\label{fig4}
\end{center}
\end{figure}

These graphs look very nice, but in fact they are deceptive. Firstly the 1st
excited state has been scaled to gap 1, and secondly the range shown
has been chosen carefully.
Extending the range of the graph for larger positive and negative
values shows that TCSA has a rather unexpected behaviour. In figure
\ref{fig3} we plot the normalised gaps for positive and negative
coupling on a logarithmic scale.
We see that the fixed points we identified earlier are not at infinite
coupling but at finite values of the coupling constant and that
apparently well organised behaviour continues beyond the ``fixed
point''.  In the case of the negative direction there appears to be a
sequence of values of the coupling constant at which the spectrum
organises itself into the $(21)$, $(12)$ and $(11)$ representations in
turn. These are exactly the sequence of flows one would expect if the
perturbing field $\phi_{13}$ in the UV transformed into the field
$\phi_{31}$ in the IR so that one passed along the normal flow from
$(31)$ through the IR fixed point $(21)$ and then proceeded in the
reverse direction towards the $(12)$ point and then again away from
$(12)$ towards the $(11)$ boundary condition. In other words this is
the same as joining two standard flows together by identifying their
mutual IR fixed point:
\[
(1\!\underbrace{1) \longleftarrow (12) \longrightarrow (2}\!%
\overbrace{1)  \longleftarrow (13) \longrightarrow (3}1)
\]
or the whole sequence of flows along the bottom of figure \ref{fig1}.
The extension of this pattern in the positive direction can be seen in
higher models, eg $M_{6,7}$ shown in figure \ref{fig4}.

In figure \ref{fig4}, the following
sequence of flows can be seen, all starting from the $(13)$ boundary condition:
\[
(1\!\underbrace{1) \longleftarrow (12) \longrightarrow (2}\!%
\overbrace{1) \longleftarrow (13) \longrightarrow (3}\!%
\underbrace{1) \longleftarrow (14) \longrightarrow \phantom{(4}}
\]
These are again exactly the flows one would expect if the perturbing field
in the UV is $\phi_{13}$  and in the IR is $\phi_{31}$ (when this
exists). 

These sequences of flows are surprising as half of the individual
flows appear to violate the $g$--theorem, namely the flows
$(21)\to(12)$ for negative $\lambda$ and $(31)\to(14)$ for positive
$\lambda$. It is worth noting that the same sequence of flows was seen
in \cite{LSS} in an apparently unrelated context.
The question is whether these flows can be understood given the
numerical nature of the TCSA and whether they affect the  
large $N$  limit of TCSA.

In the limit $N\to\infty$, the TCSA scheme is meant to approach the
TBA or NLIE 
picture in which the beta function for a single perturbation is linear
and the only fixed points are at infinity.
To find the behaviour of TCSA as $N$ varies we shall adapt the
standard method used in the field theory investigations.

\section{The TCSA renormalisation group equation}

In this section we consider a perturbation by a single field $\phi$
where the only relevant field appearing in the OPE of $\phi(x)\phi(y)$
is $\phi$ itself. This is the situation for the $\phi_{(13)}$ flows
presented in figures \ref{fig2}--\ref{fig4}.

TCSA can be thought of as standard perturbation by a field which is
projected onto states of level less than or equal to
$N$.  If this
projector is $P_N$  then we consider the perturbation by
$\lambda_N\phi_N$  where $\lambda_N$  is the effective TCSA coupling for
truncation level $N$  and $\phi_N = P_N \phi P_N$.

 We can find how $\lambda_N$  varies with
$N$  by requiring the partition
function be invariant. Rather than consider the partition function
itself directly, we can consider the operator 
\be
 {\cal P} e^{-\lambda_N\int\phi_N(x) \,dx} = 
  {\cal P} e^{-\lambda_{N+1}\int\phi_{N+1}(x) \,dx}
\;,
\ee
whose expectation value is the partition function on the strip.
Mapping this to the upper half plane, expanding this out to second
order,  and stripping off an integral we get (
with $y = 1-h$)
\begin{eqnarray}
 &
\lambda_N\phi_N(1) - \lambda_N^2 \left(\frac
 R\pi\right)^y\int_0^1\phi_N(1)\phi_N(u) \frac{du}{u^y}
\nonumber \\
=&
 \lambda_{N+1}\phi_{N+1}(1) - \lambda_{N+1}^2 \left(\frac
 R\pi\right)^y\int_0^1\phi_{N+1}(1)\phi_{N+1}(u) \frac{du}{u^y}
\;.
\end{eqnarray}
We can take the matrix element of this expression between
$\langle\phi\vert \cdots \vert 0 \rangle$
to find to second order that
\be
\lambda_{N+1}-\lambda_N
 = \left(\frac R\pi\right)^y \lambda_N^2
 \int_0^1
 \langle\phi\vert \phi(1) [P_{N+1} - P_N] \phi(u)\vert 0 \rangle
 \frac{du}{u^y}
\;.
\ee

The integrand can be identified as the coefficient of $u^{N+1}$
in the expansion of the three point function
\be
\langle\phi\vert\phi(1)\phi(u)\vert0\rangle = C (1-u)^{-h}
\;.
\ee
This coefficient is 
\be
C
\frac{\Gamma(h+N+1)}{\Gamma(h)\Gamma(N+2)}u^{N+1}
\;.
\ee
Performing the integral and taking the large $N$ limit we find 
\be
N\frac{d\lambda}{dN}
= \left(\frac{R}{N\pi}\right)^y \frac{C}{\gamma}\lambda^2
\;,
\ee
where $\gamma=\Gamma(h)$.
We can solve this exactly to find $\lambda(N)$ in terms of
$\lambda_\infty$:
\be
\lambda(N) = \frac{\lambda_\infty}{1 + (\frac{R}{N\pi})^y\frac
  C{y\gamma}\lambda_\infty}
\label{rge}
\ee
We see that
\vspace{3mm}

(1) As $N\to\infty$, $\lambda\to\lambda_\infty$

(2) As 
$\lambda_\infty\to\pm\infty\;,\;\;
\cases{ 
\ds\lambda_N\to\frac{y\gamma}C\left(\frac{N\pi}{R}\right)^y
&{for}   $\lambda C$ positive
\cr
&\cr
\lambda_N \hbox{
~diverges for a finite value of }\lambda_\infty
&{for }   $\lambda C$ negative.
\cr}$

\vspace{3mm}

Assuming that $\lambda C>0$, we see that
the IR fixed point at $\lambda_\infty=\infty$ is brought in to a
finite value. This value tends to $\infty$ as $N$ increases, so the
finite value of the fixed point is indeed an artefact of truncation
which would go away with increasing $N$.
Since this is only a first order perturbation theory calculation we
cannot expect it to provide any information about other fixed points,
for example even the closest fixed point for $\lambda C<0$ is not seen
by this calculation.

One thing to note is that if we consider
\be
\mu = \left(\frac{R}{N\pi}\right)^y\lambda
\;,
\ee
then the RG equation becomes
\be
 -N\frac{d\mu}{dN} = y\mu - \frac{C}{\gamma}\mu^2
\;.
\ee
In the limit $h$ 
tends to one this reproduces the standard beta
function of \cite{RRS} with an effective UV cutoff $a = R/(N\pi)$.
\be
\ds -N\frac{d\mu}{dN} = y\mu - {C}\mu^2
\;.
\ee
We can also see directly from here the fixed point at 
$\mu = \frac{y\gamma}C$ that
\be
\lambda_N (R/\pi)^y = \frac{y\gamma}C N^y
\;.
\ee

\section{A test of the RG equations}

\begin{figure}[hbt]
\begin{center}
\begin{tabular}{cc}
\epsfxsize=6cm
\epsfbox{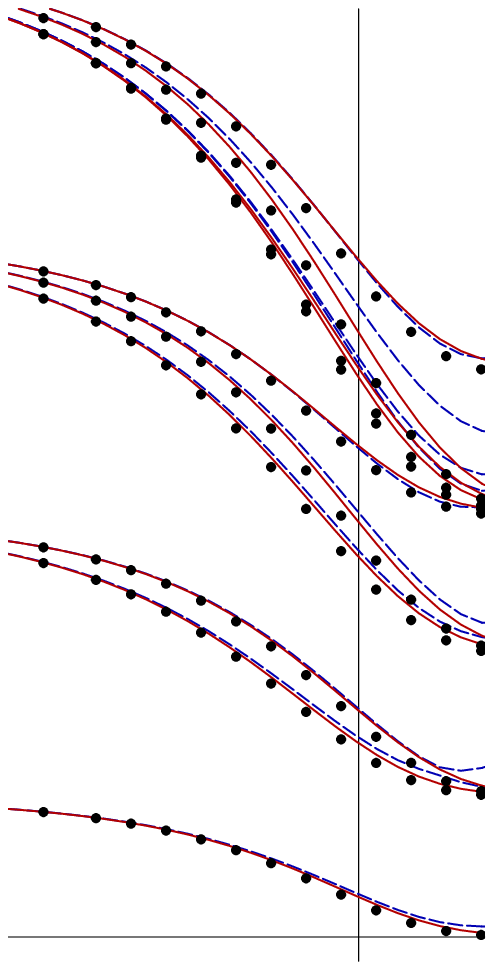}
&
\epsfxsize=6cm
\epsfbox{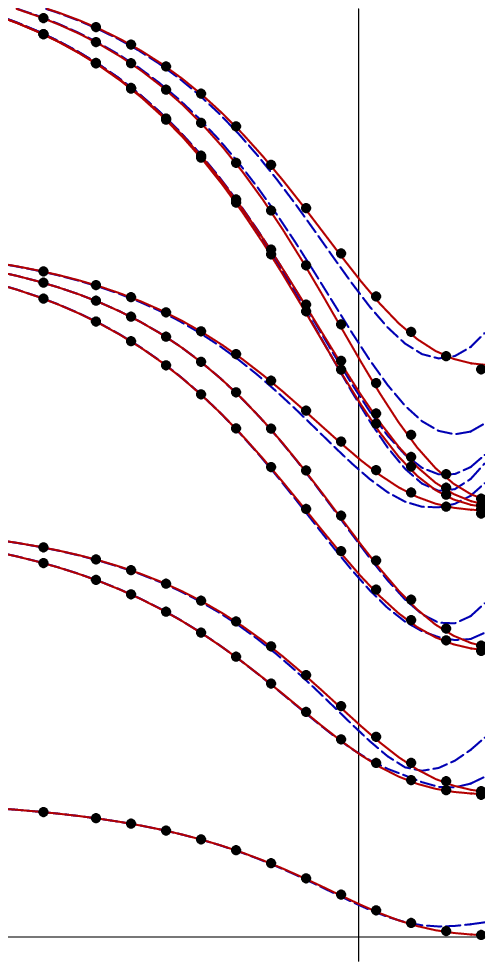}
\end{tabular}
\caption{Low-lying normalised energy gaps of the Hamiltonian in the
  tricritical Ising model on a strip with boundary conditions $(11)$
  and $(12)$ with the latter perturbed by $\phi_{(13)}$ plotted
  against the logarithm of the coupling constant. 
  On the left the TCSA data is uncorrected whereas on
  the right it is corrected using the RG equation (\ref{rge}). See text for details.
}
\label{fig5}
\end{center}
\end{figure}

We can test the RG equations by calculating the spectrum at finite $N$ with
the RG improved value of $\lambda$. 
Without RG correction the spectra will be $N$ dependent; if we use the
RG corrected value of the coupling constant then 
the spectra for different values of $N$ should agree much more
closely.
This is what we see in figure \ref{fig5} in the case%
\footnote{This is a different example to that presented in the talk
which was the boundary condition $(14)$ perturbed by the field
$\phi_{(13)}$ in $M_{6,7}$. The two examples show very
similar properties -- it has been altered to allow
inclusion of excited state TBA data which shows clearly the great
improvement in agreement resulting from the RG correction.}
of the boundary
condition $(12)$ perturbed by $\phi_{(13)}$ in the model
$M_{4,5}$. 
  The system is flowing from the boundary condition $(12)$ in the UV
  on the left to the boundary condition $(11)$ in the IR on the right.
  The (blue)
  dashed lines are at truncation level $N=5$ with 12 states and the
  (red) solid lines are at truncation level $N=16$ with 362
  states. The points are the gaps calculated using the excited state
  TBA equations. On the left the TCSA data is uncorrected whereas on
  the right it is corrected using the RG equation (\ref{rge}).
The agreement between different levels is improved by
including the RG correction and the agreement with the TBA data is
vastly improved by the RG correction.

\section{Summary and Outlook}

The fact that the TCSA method suffers from strong corrections for
finite $N$ has been known for a long time --- the corrections were
highlighted in 
\cite{YZam1} and a scaling form proposed in \cite{CLM1}. the fact that
this can bring fixed 
points in to finite values of the coupling constant was also known to
people working in the field for a long time but was regarded as an
unfortunate effect that could be removed by increasing $N$. The detailed
examination presented here was motivated by the need to obtain a
quantitative comparison between TCSA and the TBA results of
\cite{FPR1}. The detailed comparison will appear later \cite{FGPW1}.
We appear to have made a first step to understanding the finite
$N$-corrections to TCSA.

\begin{figure}[b]
\[\begin{array}{c}
\epsfxsize=0.6\hsize
\epsfbox{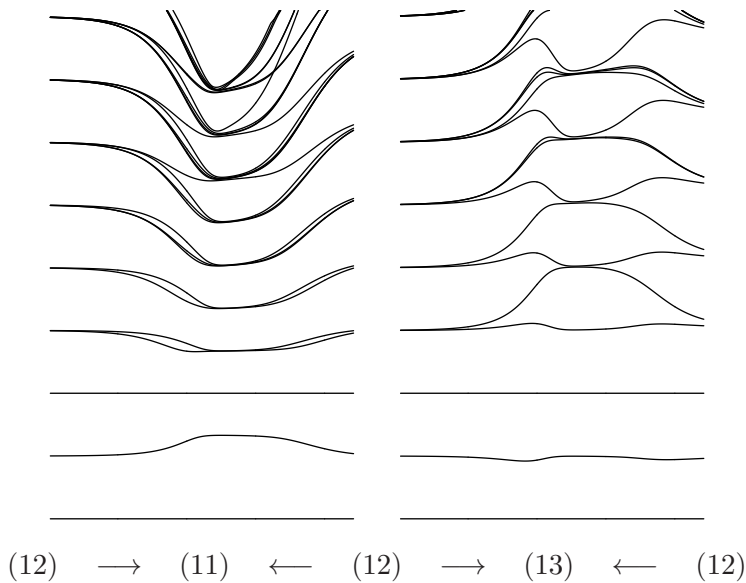}
\\
(12) 
\quad \longrightarrow \quad
(11)
\quad \longleftarrow \quad 
(12) 
\quad \longrightarrow\quad 
(13)
\quad \longleftarrow \quad
(12)
\end{array}\]
\vspace{-5mm}
\caption{The normalised gaps $\Delta'_i = 2 (E_i - E_0)/(E_2 - E_0)$
  of the Hamiltonian in the Ising model on a strip 
  with boundary conditions $(11)$ and $(12)$ perturbed by the
  boundary field $\phi_{(13)}$ in both positive and negative
  directions on a logarithmic scale - the positive on the right and
  the negative on the left. }
\label{fig6}
\end{figure}

A similar situation occurs for the Ising model but the first
correction is at third order for symmetry reasons. 
The standard TCSA truncation leads to a similar pattern of ``fixed
points'' at finite values of the TCSA coupling constant with
subsequent ``reversed'' flows as shown in figure \ref{fig6}.
However in this case there are strong
indications that the higher terms beta function depends strongly on
the type of truncation - G.~Zs.~T\'oth has constructed a different
truncation which is exactly solvable but doesn't exhibit the second
fixed point seen in ``naive'' level truncation \cite{GZsT1,TW1}.

\begin{figure}[h]
\begin{center}
\epsfxsize=0.5\hsize
\epsfbox{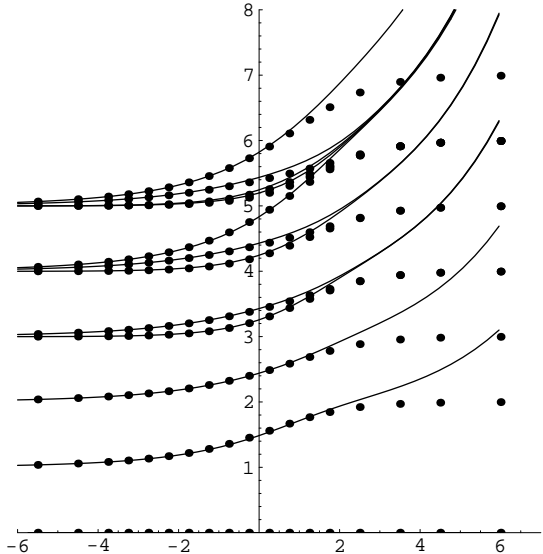}
\caption{TBA and TCSA results for the energy gaps in the tricriical
  Ising model on a strip with $(11)$ and $(12)$ boundary conditions
  perturbed by the field $\phi_{1,3}$ flowing to the boundary
  conditiion $(1,1)$ on a logarithmic scale.}
\label{fig7}
\end{center}
\end{figure}

There is however another important point which remains to be
understood quantitatively and which has also been concealed in the
results shown here. That is the need for an overall rescaling of the
Hamiltonian, or an effective change in the strip width. In each of
figures \ref{fig1} to \ref{fig5}, it is the normalised energy gaps
which have been plotted. 
As a final plot, in figure \ref{fig7} we show an example of
unnormalised energy gaps. A quantitative understanding of this
rescaling still remains elusive.

\section{Acknowledgments}

This talk was presented by GW at the workshop {\em ``Integrable Models
and Applications: from Strings to Condensed Matter''} based on results
obtained over an extended period with the other authors and
He would like to thank them for a very interesting collaboration and 
Patrick Dorey, Andreas Recknagel, Daniel Roggenkamp, Ingo Runkel,
Volker Schomerus, Gabor Takacs and Roberto Tateo for useful discussions.
He would also like to thank the organisers of the meeting for the chance
to present these results for the first time.
This work has been supported at various times by
the EU network ``EUCLID'', contract number HPRN-CT-2002-00325
and PPARC rolling grant PP/C5071745/1.

\end{document}